%% file: camera_ready.tex
\documentclass[letterpaper]{article} 
\usepackage{aaai24}  
\usepackage{times}  
\usepackage{helvet}  
\usepackage{courier}  
\usepackage[hyphens]{url}  
\usepackage{graphicx} 
\usepackage{natbib}  
\usepackage{caption} 
\usepackage{algorithm}
\usepackage{algorithmic}
\usepackage{adjustbox}

\usepackage{multirow}
\usepackage{booktabs}

\usepackage{amssymb}
\usepackage{amsthm}

\newtheorem{theorem}{Theorem}[section]

\newtheorem{lemma}[theorem]{Lemma}

\newcounter{subassumption}[assumption]

\makeatletter
\renewcommand{\p@subassumption}{\theassumption}
\makeatother

\usepackage{newfloat}
\usepackage{listings}
\DeclareCaptionStyle{ruled}{labelfont=normalfont,labelsep=colon,strut=off} 
\floatstyle{ruled}
\newfloat{listing}{tb}{lst}{}
\floatname{listing}{Listing}
\title{CREAD: A Classification-Restoration Framework with Error Adaptive Discretization for Watch Time Prediction in Video Recommender Systems}  
\author {
    Jie Sun,
    Zhaoying Ding,
    Xiaoshuang Chen,
    Qi Chen,
    Yincheng Wang,
    Kaiqiao Zhan,
    Ben Wang\footnote{Corresponding Author}
}
\affiliations {
    Kuaishou Technology, Beijing, China \\
     jie.sun327@gmail.com,\{dingzhaoying, chenxiaoshuang, chenqi11, wangyincheng,
    zhankaiqiao, wangben\}@kuaishou.com
}

\usepackage{bibentry}

\usepackage{xcolor}
\input{math_commands}

\begin{document}

\maketitle

\begin{abstract}
The watch time is a significant indicator of user satisfaction in video recommender systems.
However, the prediction of watch time as a target variable is often hindered by its highly imbalanced distribution with a scarcity of observations for larger target values and over-populated samples for small values. State-of-the-art watch time prediction models discretize the continuous watch time into a set of buckets in order to consider the distribution of watch time. However, it is highly uninvestigated how these discrete buckets should be created from the continuous watch time distribution, and existing discretization approaches suffer from either a large learning error or a large restoration error. To address this challenge, we propose a Classification-Restoration framework with Error-Adaptive-Discretization (CREAD) to accurately predict the watch time. The proposed framework contains a discretization module, a classification module, and a restoration module. It predicts the watch time through multiple classification problems. The discretization process is a key contribution of the CREAD framework. We theoretically analyze the impacts of the discretization on the learning error and the restoration error, and then propose the error-adaptive discretization (EAD) technique to better balance the two errors, which achieves better performance over traditional discretization approaches.
We conduct detailed offline evaluations on a public dataset and an industrial dataset, both showing performance gains through the proposed approach. Moreover, We have fully launched our framework to Kwai App, an online video platform, which resulted in a significant increase in users' video watch time by 0.29$\%$ through A/B testing. These results highlight the effectiveness of the CREAD framework in watch time prediction in video recommender systems.

\end{abstract}
\section{Introduction}

Recommender systems have seen great success in matching users with items they are interested in
\cite{herlocker2004evaluating}. 
One of the most popular applications is short-video social media \cite{tang2017popularity, wu2018beyond} such as TikTok and Instagram Reels, where short videos appear on the user's screens without any active operation such as clicking. Consequently, traditional metrics such as click-through rates are non-applicable anymore. Intuitively, watch time becomes a crucial metric for measuring user engagement \cite{covington2016deep}. To ensure optimal user experience, it is essential to predict watch time in online recommender systems accurately. By doing so, these platforms can gain a better understanding of user preferences and make personalized video recommendations tailored to their interests.
A large body of research \cite{zhan2022deconfounding,gong2022real,lin2022feature,wang2022make,cai2023reinforcing, zhao2023disentangled} has been devoted to developing neural network models, and significantly improves the prediction accuracy over traditional regression methods.


\begin{figure}[t]
\centering
\includegraphics[width=0.8\columnwidth]{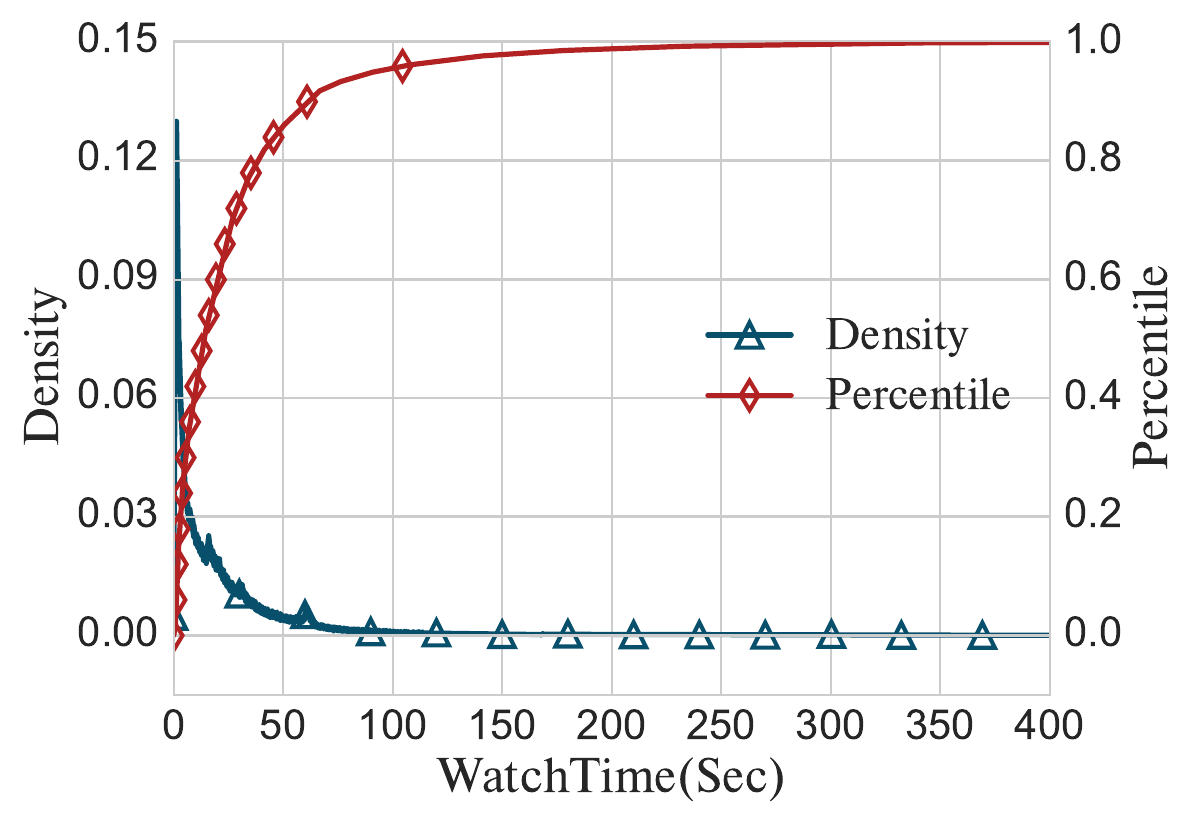}
\caption{Probability density plot of watch time.}
\label{fig:watchtime_distribution}
\end{figure}



The prediction of watch time poses a regression problem due to the continuous and wide-ranging nature of users' watch time values, which amplifies susceptibility to outliers and potential prediction skew. The distribution of watch time for short videos, as depicted in Figure \ref{fig:watchtime_distribution}, is right-skewed, featuring substantial concentrations within limited time frames: $30$\% of views within $3$ seconds and $80$\% within $32$ seconds. This distribution presents a challenge for earlier watch time prediction attempts \cite{zhan2022deconfounding, covington2016deep}, which neglected the long-tailed nature of regression problems, thus yielding suboptimal results for tail instances. Moreover, in recommendation systems, ordinal relationships among predictions play a key role, with watch time as a metric for video comparison, highlighting the importance of ordinal ranking. However, standard regression losses like $\ell_1$ and $\ell_2$ do not consider ranking comparisons, focusing solely on discrepancy magnitude. Consequently, maintaining prediction accuracy and ranking efficacy within real-time recommender systems amid imbalanced continuous label distributions poses great challenges. 


To address this, we introduce an effective classification-restoration-based framework for learning from imbalanced continuous targets in real-world contexts. The framework contains three key components: an effective discretization to convert continuous labels into ordinal intervals, a classification module training multiple binary classifiers across these segments to ensure ranking accuracy and a restoration module for predicting the watch time from the predictions of these classifiers. 

A challenge with this approach is the ambiguity in creating discrete classes from the continuous distribution. This entails addressing two error types: the \emph{learning error}, tied to sample bucket count, and the \emph{restoration error}, which impacts watch time estimation from discretized predictions. Balancing these errors proves complex; narrow bucket intervals lower the learning error by reducing sample probability, while wider intervals diminish information and elevate the restoration error. We examine discretization's effect on learning and restoration errors and propose an error-adaptive discretization (EAD) method that harmonizes these errors within practical distributions.

Our comprehensive framework, named Classification-Restoration with EAD (CREAD), offers a versatile extension applicable to existing learning methods, such as D2Q.
In summary, our main contributions are:
\begin{itemize}
\item We propose a general classification-restoration framework for learning the ordinal information of watch time, together with the training algorithm to reduce the classification and restoration error.
\item We analyze the error bounds introduced by discretization, and propose a novel discretization approach to balance between the learning error and the restoration error, according to the realistic dataset distribution.

\item 
Both offline and online experiments with realistic large-scale datasets show that our framework achieves competitive results compared with state-of-the-art methods. 
\end{itemize}

\section{Related Works}
\label{sec: relatedwork}
\subsection{Watch Time Prediction}
Watch time prediction is the task of predicting the watch time of a user, given the user profile, interactive history, and a series of candidate short videos. 
Value Regression (VR) directly predicts the absolute value of watch time, where the learned function's accuracy is evaluated by mean squared error. 
\citet{covington2016deep} incorporates watch time as a sample weight in the logistic regression of (WLR) impressive videos, transforming the direct regression of watch time into learned odds of video click-through rate. However, this assumption holds only when the impression rate is low and is not applicable in short video settings where impressed videos are played automatically. Recently,  \citet{zhan2022deconfounding} studied duration bias in watch time prediction for video recommendation (D2Q), removing undesired bias by binning data based on duration. Although their equal-frequency-based methods removed the bias, they ignored the impact of the imbalanced distribution of watch time on long-tailed samples, resulting in lower accuracy compared to head samples. Although effective, they did not leverage the additional bucket-wise information lost in the discretization process, while we impose an error-adaptive framework in the modeling process.

\subsection{\label{sec:RvC_related}Regression by Classification} 
Recently, there is a trend to pose the regression problem as a set of classification problems, which has a significant improvement on direct regression. 
The first related work is Ordinal Regression (OR). It is initially used for classification problems where the dependent variable exhibits a relative ordering and later extended to multiple fields, including age estimation  \cite{niu2016ordinal, beckham2017unimodal} and depth estimation \cite{diaz2019soft}. 
A vital problem of OR is the ambiguity in how the discrete classes should be created from the distribution. 
Most works use fixed standard methods such as equal-width or equal-frequency discretization to divide the continuous variables, while others manually choose multiple thresholds \cite{crammer2001pranking, shashua2002ranking} as hyperparameters. 
As will be analyzed in Sec. \ref{sec: error_analysis}, these methods introduce a large discretization error, especially when the data follows an imbalanced distribution. In contrast, our approach alleviates the problem by proposing an adaptive discretization method that minimizes the total error. 

\section{Method\label{sec: overview}}
\textbf{Notations.} Let $\{(\vx_i, y_i)\}^N_{i=1}$  be the training set, where $y_i \in \gY \subset \bbR^+$ is the corresponding ground truth watch time, and $\vx_i \in \bbR^d$ denotes the $i$-th input, including user-related features (such as demographics characteristics and browsing histories) and video-related features (such as the tags and forwarding counts). 
Without loss of generality, we assume the domain of target variables is bounded by $T_{\max} \in \bbR^+$.
We use $M-1$ thresholds $\{t_m\}_{m=1}^{M-1}$ to divide the domain into $M$ discrete buckets $\gD \triangleq \{d_m\}_{m=1}^M$, where the $m$-th bucket $d_m = [t_{m-1}, t_m)$ for $m=1, \cdots, M$, and $t_0 = 0$, $t_M = T_{\max}$.
Let $\hat{y}_i$ denote the predicted watch time of $\vx_i$. 
Let $\bone(\cdot)$ denote the indicator function. 
For simplicity, we omit the subscript $i$ when we do not specify a certain sample. 



\subsection{Overall Framework}
\begin{figure}[t]
	\centering
\includegraphics[width=0.9\columnwidth,trim=0 100 0 0]{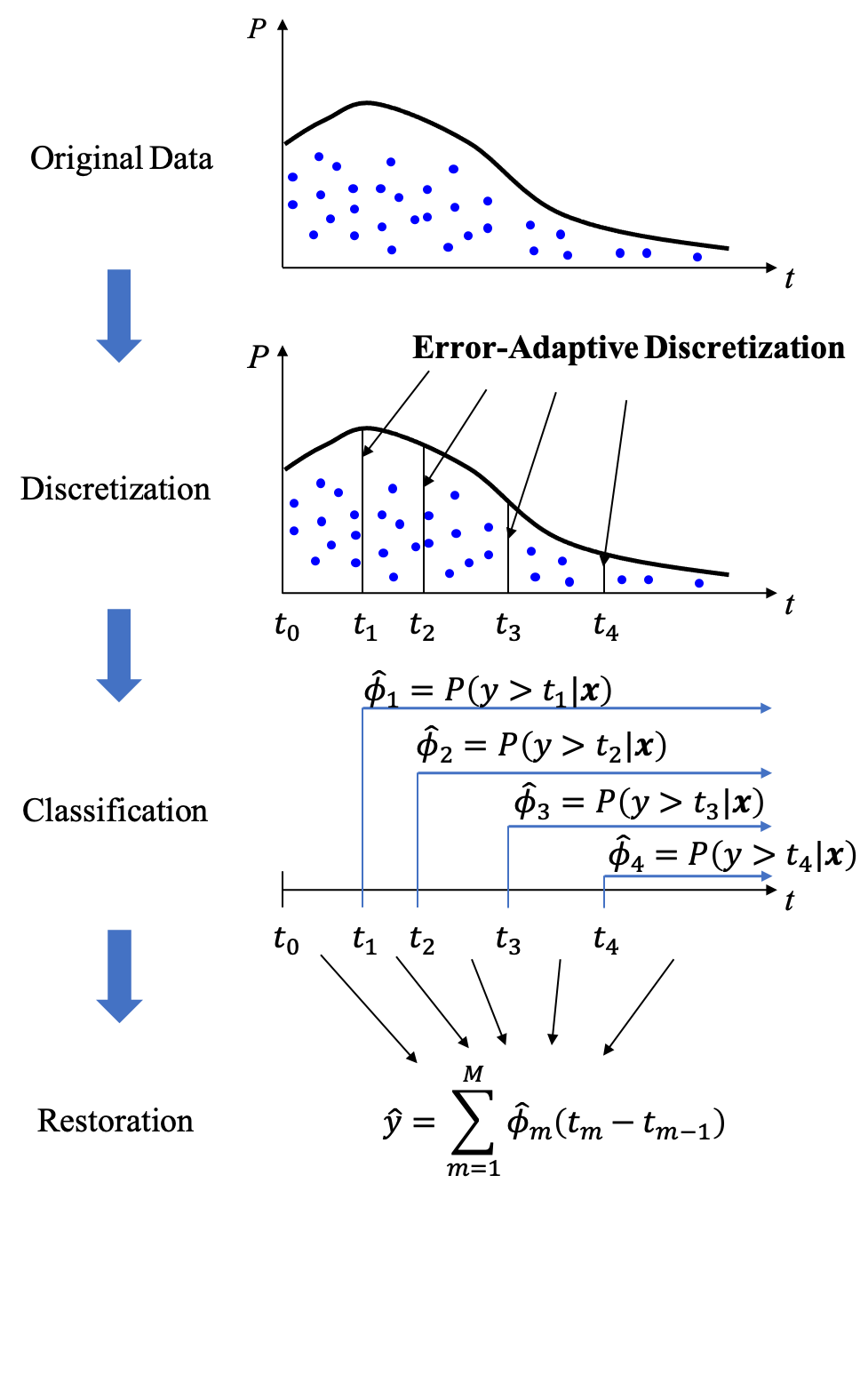}
\caption{The CREAD Framework. }
\label{fig:framework}
\end{figure}

The CREAD Framework, as shown in Figure \ref{fig:framework}, includes three modules, \ie discretization, classification, and restoration. In the following, we explain the design of each component.

\subsubsection{Discretization}
This module is a preprocess module independent of the training and evaluation process. It obtains thresholds $\{t_m\}_{m=1}^{M-1}$ according to the data distribution and break the target domain $\gY$ into $M$ non-overlapping buckets $\gD \triangleq \{d_m = [t_{m-1}, t_m)\}_{m=1}^M$. These buckets are used to transform the watch time $y$ into $m$ discretized labels:
\begin{equation} \label{eq:y-m}
    y_m = \bone(y > t_m).
\end{equation}

The discretization strategy is crucial for prediction accuracy, and we will discuss it in detail in Sec. \ref{sec:error-adaptive-discretization}.

\subsubsection{Classification} 
\label{sec:adld}
$M$ classifiers are trained to predict whether the watch time $y$ is larger than the $m$-th threshold $t_m$, \ie $y_m$ in \eqref{eq:y-m}, and output a series of probabilities:
\begin{equation} \label{eq:hat-phi-m}
    \hat{\phi}_{m}(\vx_i; \Theta_m) =  P(y>t_m|\vx_i),~ 1 \leq i \leq N.
\end{equation}
The classifiers are neural networks with learnable parameters $\Theta_m$. We introduce how to train the models in Sec. \ref{sec:lossfunc}.




\subsubsection{Restoration}
Given the $\{\hat{\phi}_m\}_{m=1}^M$, we are able to restore the predicted watch time. The restoration is based on the following fact of expectation:
\begin{equation}
\begin{aligned} 
\bbE(y|\vx_i) 
& = \int_{t=0}^{t_M} tP(y=t|\vx_i) dt \\
& = \int_{t=0}^{t_M} P(y > t|\vx_i)dt \\
& \approx \sum_{m=1}^{M} P(y > t_m|\vx_i)\left(t_m-t_{m-1}\right).
\end{aligned}
\end{equation}

By the definition of $\hat{\phi}_m$ in \eqref{eq:hat-phi-m}, we can reconstruct the predicted watch time from these discretized predictions $\hat{\phi}_m$:
\begin{equation} \label{eq:restoration}
\hat{y} = \sum_{m=1}^M \hat{\phi}_{m}\left(t_{m}-t_{m-1}\right).
\end{equation}

\subsection{Model Training\label{sec:lossfunc}}
Here, we provide the loss function in the training of $M$ classifiers. The loss functions contain three parts, of which the first is the \textbf{classification loss} via standard cross-entropy.
\begin{align} \label{eq:discrte_loss}
\gL_{\text{ce}} = \sum_{m=1}^{M}-y_{m}\log(\hat{\phi}_{m}) -(1-y_{m})\log(1-\hat{\phi}_{m}).
\end{align}



The second is the  \textbf{restoration loss} to reduce the error of the reconstructed watch time in \eqref{eq:restoration}: 
\begin{align}
\label{eq:acc_loss}
    \gL_{\text{restore}} &= \ell(\hat{y}, y),
\end{align}
where $\ell$ is a loss function to measure the deviation from $\hat{y}$ to $y$. We find it beneficial to apply Huber loss\cite{huber1992robust} for $\gL_{\text{restore}}$, which will be detailed analyzed in Sec. \ref{sec:ablation}. 

The third is a \textbf{regularization term by ordinal prior}.
By definition, the output of the $M$ classifiers $\{\phi_m\}_{m=1}^M$ has a prior by definition, \ie $\hat{\phi}_{m}$ monotonically decreases as $m$ grows. Consequently, we impose the prior into the proposed framework by minimizing the following regularization:
\begin{align}
\gL_{\text{ord}}= \sum_{m=1}^{M-1}  \max(\hat{\phi}_{m+1} - \hat{\phi}_{m}, 0).
\end{align}

In summary, the final optimization objective is
\begin{align}\label{eq:final_loss}
    \gL_{\text{CREAD}} = \lambda _{\text{ce}} \gL_{\text{ce}}  + \lambda _{\text{restore}} \gL_{\text{restore}}
    + \lambda _{\text{ord}} \gL_{\text{ord}},
\end{align}
where $\lambda _{\text{ce}}$, $\lambda _{\text{restore}}$, and $\lambda _{\text{ord}}$ are hyperparameters. 


\subsection{Challenges of Discretization}
\begin{figure}[t]
    \centering
\includegraphics[width=1.0\columnwidth]{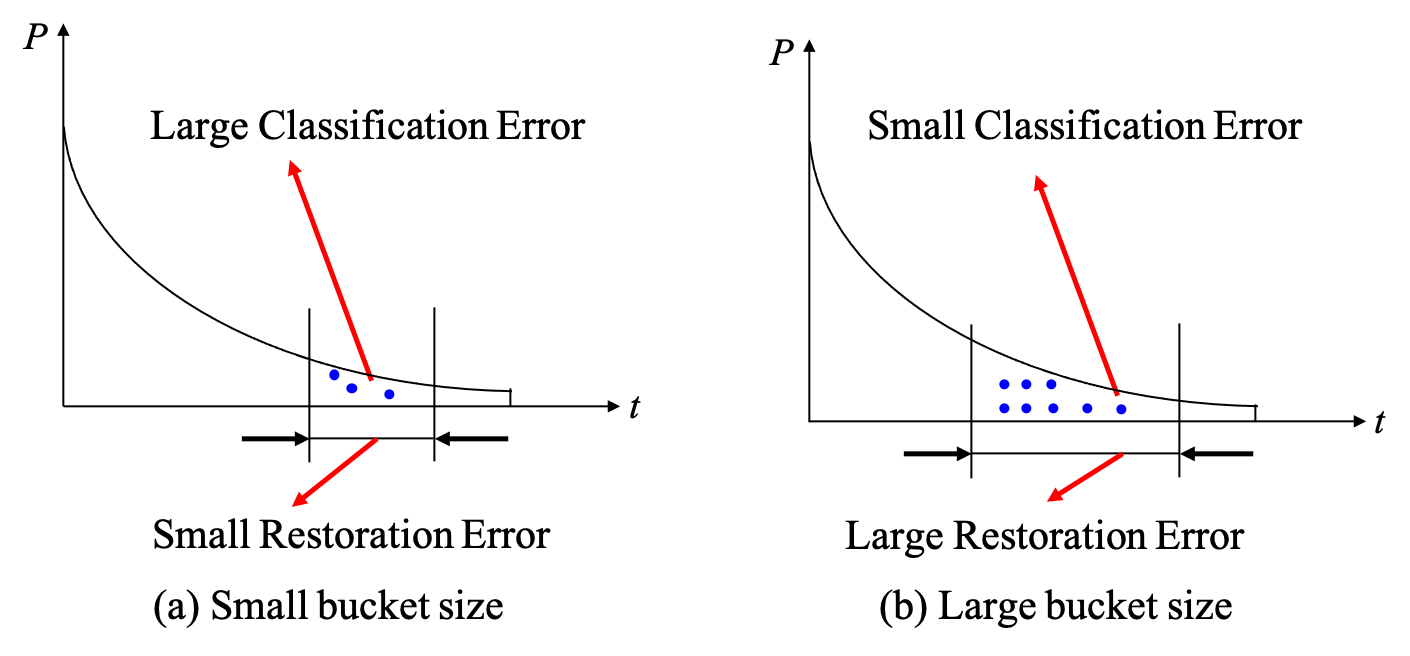}
    \caption{Two kinds of errors in the discretization.}
    \label{fig:2-errors}
\end{figure}

In the CREAD framework, a key module is the discretization module, and the method of discretization largely affects the final prediction accuracy. 
As shown in Figure \ref{fig:2-errors}, the discretization introduces two kinds of errors:
\begin{itemize}
    \item \textbf{Learning Error}: Since the number of instances in each bucket is finite, the $M$ classifiers cannot be infinitely accurate. As we increase the number of buckets $M$, the number of instances that fall in each bucket decreases, thus limiting the classification performance. 
    \item \textbf{Restoration Error}: The restoration in \eqref{eq:restoration} is an approximate function of the expectation, omitting the detailed probability density in each bucket $[t_{m-1}, t_m]$, which will also introduce errors.
\end{itemize}

Unfortunately, these two errors cannot be simultaneously reduced. To reduce the learning error, a larger bucket width is needed, leading to a larger restoration error (see Figure \ref{fig:2-errors}). Existing approaches usually use equal-width or equal-frequency methods \cite{gai2017learning} to heuristically set the discretization set $\mathcal{D}$. We will show in Sec. \ref{sec: error_analysis}  that both equal-width and equal-frequency methods cannot balance the two errors well, and propose our EAD approach.

\section{Balance the Errors in Discretization}\label{sec: error_analysis}
This section aims to balance the errors introduced in the discretization process. We first provide a theoretical analysis of the error bounds of the learning error and the restoration error introduced in the discretization process, and then propose the EAD approach to effectively balance the two errors.

\subsection{Decomposition of Errors from Discretization}
Assume the training dataset $\{(\vx_i, y_i)\}^N_{i=1} \sim \mu(\vx, y)
= \mu(\vx)\mu(y|\vx)
$ is \emph{i.i.d.} distributed. 
Let $p_m(\vx) = P(y \in d_m | \vx)$ denote the probability that the label $y$ belongs to the $m$-th bucket $d_m$ given $\vx$.
Let $v_m(\vx) = \bbE(y|\vx, y\in d_m)$ be the expectation of the watch time of sample $\vx$, given that it belongs to the $m$-th bucket. 
Let $w_m = \bbE_{\vx\sim\mu(\bx)}v_m(\vx)$ denote the expectation of watch time in the interval $d_m$.

We add the hat superscript to denote the predicted value, \eg $\hat{p}_m(\vx)$ as the prediction of $p_m(\vx)$ and $\hat{w}_m$ as the prediction of $w_m$. Then we can formulate the watch time as $\hat{y} = \sum_m\hat{p}_m(\vx)\hat{w}_m$. Notice that such form is equivalent to the cumulative form in \eqref{eq:restoration}, where $\hat{p}_m = \hat{\phi}_m - \hat{\phi}_{m-1}$. 

Now we aim to estimate the error bound between the predicted watch time $\hat{y}$ and the ground truth $y$. To achieve this, we first provide a decomposition of the errors:

\begin{lemma} \label{lemma:error-decomposition}
Assume that $\hat{p}_m(\vx)$ and $\hat{w}_m$ are unbiased estimations of $p_m(\vx)$ and $w_m$, respectively, we have 
\begin{equation}
\bbE\left(\hat{y} - y\right)^2 = V_p + V_w + V_b + V_y,
\end{equation}
where
\begin{align}
V_p &= \bbE_{\vx}\bbE_{\hat{p}}\bbE_{\hat{w}}\left[\sum_m\left(\hat{p}_m(\vx) - p_m(\vx)\right)\hat{w}_m\right]^2, \\
V_w &= 
\bbE_{\vx}\bbE_{\hat{w}}\left[\sum_mp_m(\vx)\left(\hat{w}_m - w_m\right)\right]^2, \\
V_b &= \bbE_{\vx}\left[\sum_mp_m(\vx)\left(w_m - v_m(\vx)\right)\right]^2, \\
V_y &= \bbE_{\vx, y}\left[\sum_mp_m(\vx)v_m(\vx) - y\right]^2.
\end{align}
\end{lemma}

Please refer to Appendix. \ref{sec: proof} for the detailed proof. Intuitively,
$V_p$ is determined by the learning error of the probability that $y$ falls into each bucket, i.e., $p_m(\vx)$. $V_w$ describes the impacts of learning error on the representing value $\hat{w}_m$, $V_b$ is the error caused by reconstructing the watch time from discretization, and $V_y$ is the intrinsic variance of the watch time $y$.

The two prediction errors $V_p$ and $V_w$ are affected by the error of the learning algorithm. Therefore, these two error terms correspond to the learning error. In contrast, $V_b$ corresponds to the restoration error independent of the concrete learning algorithms. Finally, $V_y$ is unrelated to either the learning or the discretization processes, and will not be discussed later.

\subsection{Error Bounds of Discretization} \label{sec:error-bound}
This section analyzes how the discretization process affects the error bounds.  For ease of theoretical analysis, we only consider the case of tabular input and assume that $\mu(\vx, y)$ is sufficiently smooth. But we emphasize that the discretization approach inspired by the theoretical analysis will also be effective in real-world settings.

\begin{theorem} \label{thm:error-bound}
    Assume that the input $\vx$ is sampled from a finite set $\mathcal{X}$. Moreover, assume $\hat{p}_m(\vx), \vx\in\mathcal{X}$ and $\hat{w}_m$ are obtained from maximum likelihood estimation. Moreover, assume $\mu(\vx, y)$ is with a bounded second partial derivative. Then we have
\begin{align}
    V_p & \leq \overline{V}_p \triangleq \frac{C_p\left|\mathcal{X}\right|}{N}\cdot A_p(\mathcal{D}), \\
    V_w & \leq\overline{V}_w \triangleq \frac{C_w}{N}\cdot A_w(\mathcal{D}), \\
    V_b & \leq \overline{V}_b \triangleq C_b \cdot A_b(\mathcal{D}),
\end{align}
where $C_p,C_w$ and $C_b$ are constants independent of the discretization $\mathcal{D}$, and $A_p$, $A_w$, $A_p$ are functions of $\mathcal{D}$:
\begin{align}
    A_p(\mathcal{D}) =&M\bbE_{y\sim\Psi}y^2, \label{eq:A-p} \\
    \begin{split}
        A_w(\mathcal{D}) =&\sum_{m\in\mathcal{M}} \left[\Psi(t_m) - \Psi(t_{m-1})\right]^2  \\
        &\cdot\sum_{m\in\mathcal{M}}\frac{\left(t_m-t_{m-1}\right)^2}{\Psi(t_m)-\Psi(t_{m-1})},  \label{eq:A-w}
    \end{split}\\
    \begin{split}
        A_b(\mathcal{D}) =& \sum_{m\in\mathcal{M}} \left[\Psi(t_m) - \Psi(t_{m-1})\right]^2 \\
        &\cdot\sum_{m\in\mathcal{M}}\left(t_m-t_{m-1}\right)^2, \label{eq:A-b}
    \end{split}
\end{align}
where $\Psi$ is the CDF of watch time $y$:
\begin{equation}
    \Psi\left(t\right) \triangleq P\left\{y \leq t\right\} = \bbE_\vx \int_0^t\mu(y|\vx)dy.
\end{equation}

\end{theorem}
\begin{proof}
See Appendix B.
\end{proof}


Therefore, we find that the prediction error is bounded by several functions $A_p,A_w$ and $A_b$, which only depend on the watch time distribution $\Psi$ and the discretization $\mathcal{D}$.
Now we provide some intuitive explanations.


\subsubsection{Discussion}
\label{sec:dis}
What are the impacts of different discretization methods on each error term? Here we mainly discuss the error terms $A_w$ and $A_b$, since they depend on the division points $\{t_m\}_{m=1}^M$ of $\mathcal{D}$. We consider a truncated exponential distribution on $[0,1]$, \ie $\Psi(t) = (1-e^{-5t}) / (1-e^{-5})$, discretized by $10$ buckets. Table \ref{tab:4-error-term} shows the different terms on the equal-width and equal-frequency discretization. It is shown that the equal-width method leads to lower $A_b$, while the equal-frequency method leads to lower $A_w$. This result has a very intuitive explanation, which shows the meaning of $A_b$ and $A_w$:
\begin{itemize}
    \item \textbf{Learning Error}: The learning error is shown by $A_w$, which is affected by the number of samples in each bucket. Specifically, there exists a $\Psi(t_m)-\Psi(t_{m-1})$ term in the denominator of $A_w$. Therefore, if there are few samples in some buckets, the corresponding probability $\Psi(t_m)-\Psi(t_{m-1})$ will be small, leading to a large error.
    \item \textbf{Restoration Error}: $A_b$ shows the restoration error bound. It contains a $t_m-t_{m-1}$ term as a multiplier. When $t_m-t_{m-1}$ is large for some $m$, the error term will increase, which is consistent with the intuition that larger bucket width will lead to a larger restoration error.
\end{itemize}

Now we discuss again the dilemma of errors introduced by discretization: \textbf{the learning error and the restoration error usually contradict}. If we are about to decrease the learning error, we need to increase the samples in each bucket, but a large bucket width leads to a large restoration error. Formally, the bucket probability $\Psi(t_m)-\Psi(t_{m-1})$ is usually positively related to the bucket width $t_m - t_{m-1}$. Given the abovementioned discussions, we are about to provide the EAD approach to balance the two kinds of errors.



\begin{table}[t]
\centering
\begin{tabular}{ccc} 
\toprule
Method & $A_w(\mathcal{D})$ & $A_b(\mathcal{D})$ \\
\midrule
Equal-Width  & 1.42 & \textbf{0.025} \\
Equal-Freq & \textbf{0.34} & 0.034  \\
\bottomrule
\end{tabular}
\caption{ Different Error Terms on Different Methods.}
\label{tab:4-error-term}
\end{table}




\subsection{EAD Approach} \label{sec:error-adaptive-discretization}
\begin{figure}
\centering
\includegraphics[width=1.0\columnwidth]{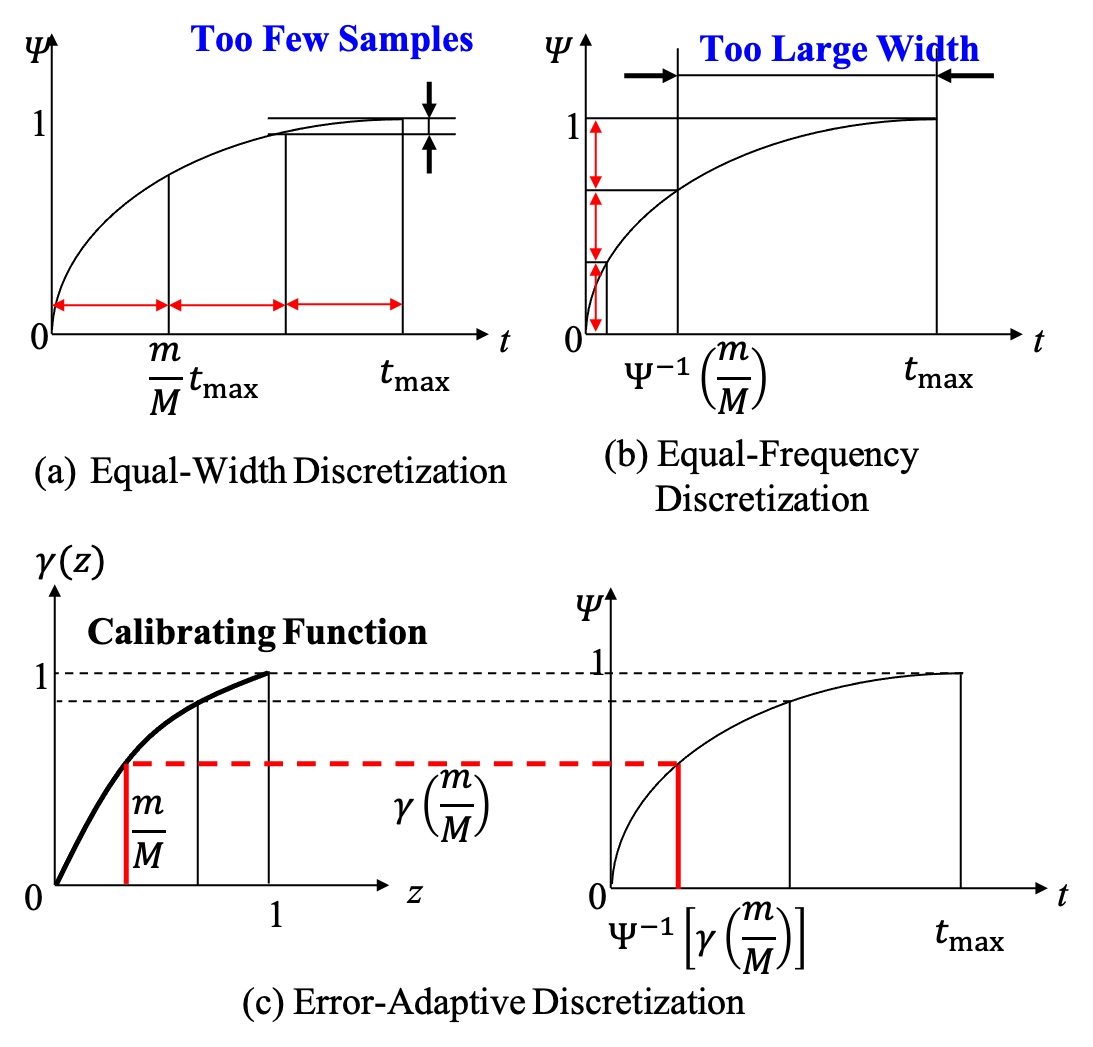}
\caption{EAD compared with traditional methods.}
\label{fig:discretization_comparison}
\end{figure}

Here we mainly discuss the discretization strategy $\mathcal{D}$ given the number of buckets $M$. We do not discuss the choice of $M$ since it is a single variable that can be regarded as a hyperparameter. According to Sec. \ref{sec:dis}, the discretization strategy $\mathcal{D}$ needs to balance the learning error $A_w$ and the restoration error $A_b$. Thus, the discretization strategy of EAD minimizes the following objective:
\begin{equation} \label{eq:bucket-opt}
    \min_{\mathcal{D}} J(\mathcal{D}) =  A_w(\mathcal{D}) + \beta A_b(\mathcal{D}),
\end{equation}
where $\beta$ is determined by $C_w$, $C_b$, and $N$ according to Theorem \ref{thm:error-bound}. Since $C_w$ and $C_b$ depend on the characteristics of the datasets, which cannot be obtained theoretically, we regard $\beta$ as a hyperparameter.

\eqref{eq:bucket-opt} is an optimization problem with dimension $M-1$. It is challenging to find the optimal discretization strategy
since $M$ is usually tens or hundreds. Here we propose a more lightweight method. We begin by expressing the equal-width and equal-frequency discretization methods formally. Specifically, the equal-width method writes:
\begin{align} 
\label{eq:equal_width}
t_m = \frac{m}{M}T_{\text{max}},
\end{align}
which guarantees a fixed $\Delta t_m = T_{\text{max}} / M$, but leads to too small $\Delta\Psi(t_m)$ in the long-tailed buckets (see Figure \ref{fig:discretization_comparison}(a)). In contrast, the equal-frequency method writes:
\begin{align} 
\label{eq:equal_frequency}
t_m = \Psi^{-1}\left(\frac{m}{M}\right),
\end{align}
which guarantees a fixed $\Delta\Psi(t_m) = 1 / M$, but leads to too large $\Delta t_m$ in the long-tailed buckets (see Figure \ref{fig:discretization_comparison}(b)).

The key is to rewrite \eqref{eq:equal_width}$\sim$(\ref{eq:equal_frequency}) by:
\begin{align} 
\label{eq:adaptive_label_discretization}
t_m = \Psi^{-1}\left[\gamma\left(\frac{m}{M}\right)\right],
\end{align}
where $\gamma$ is a calibrating function $\gamma: [0,1]\to [0,1]$, with $\gamma(0)=0, \gamma(1)=1$. Note that when $\gamma(z) = \Psi(zT_{\text{max}})$, we obtain \eqref{eq:equal_width}; and when $\gamma(z) = z$, we obtain \eqref{eq:equal_frequency}. 

Therefore, \eqref{eq:adaptive_label_discretization} contains two extreme cases of the discretization methods. This inspires us that by properly choosing the calibrating function $\gamma$ between the real distribution $\Psi$ and the uniform distribution, we can obtain any intermediate bucketing strategies. Figure \ref{fig:discretization_comparison}(c) shows the impacts of the calibrating function $\gamma$ on the discretization process.

According to the abovementioned discussions, we set $\gamma$ to be in a group of functions $\gamma(\cdot;\alpha)$, parameterized by $\alpha$, then it is possible to find the optimal $\gamma$ under the optimization problem in \eqref{eq:bucket-opt} by grid search.

\begin{figure}
\centering
\includegraphics[width=0.8\columnwidth]{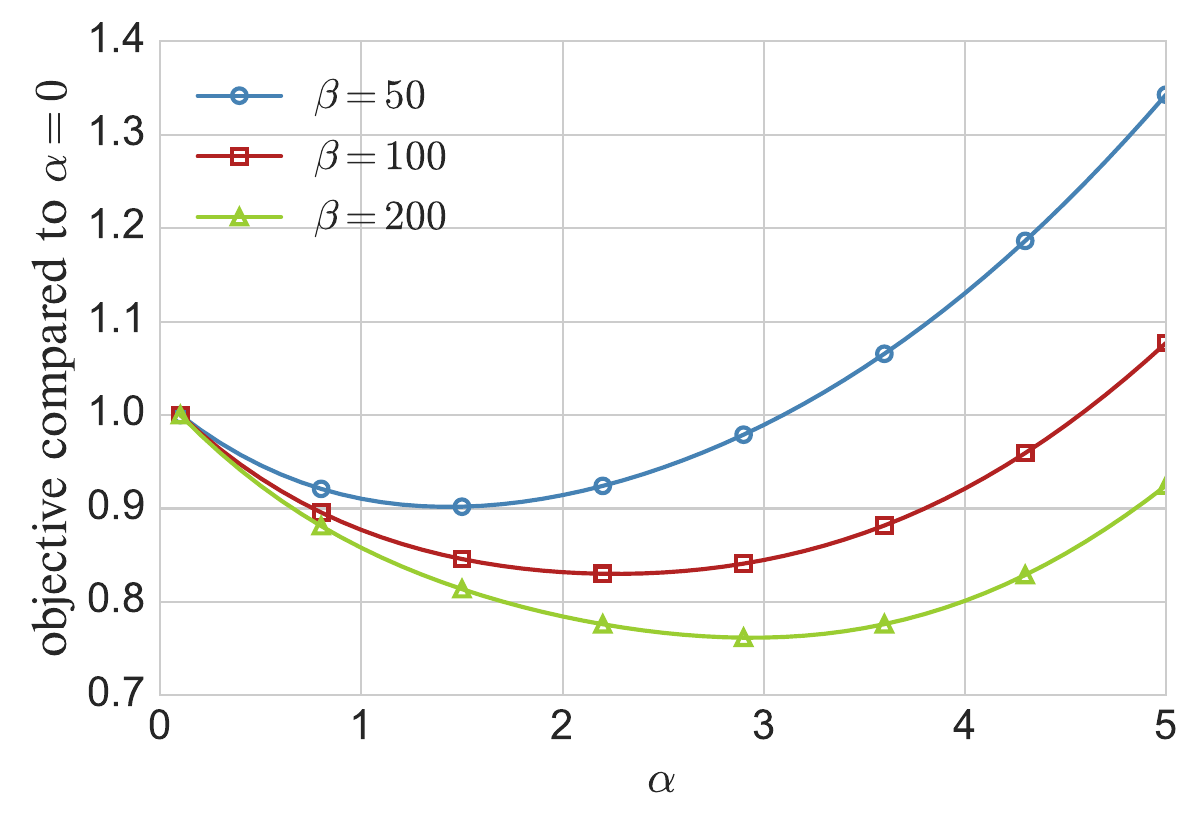}
\caption{Error terms in different hyperparameters.}
\label{fig:discretization}
\end{figure}

As an illustration, we use the same settings as Discussion 2 in Sec. \ref{sec:error-bound}, and set $\gamma$ to be $\gamma(z;\alpha) = \left(1-e^{-\alpha z}\right)/\left(1-e^{-\alpha}\right)$ with a parameter $\alpha$ from 0 to 5, and set $\beta = 50/100/200$. Specifically, $\alpha=0$ degrades to the equal-frequency method, while $\alpha=5$ degrades to the equal-width method. Figure \ref{fig:discretization} shows the objective function $J(\mathcal{D})$ (\eqref{eq:bucket-opt}) under $\alpha$ and $\beta$, which shows
\begin{itemize}
    \item The hyperparameter $\beta$ allows us to balance the learning error and the restoration error flexibly.
    \item By comparing the optimal $\alpha$ with the equal-frequency method ($\alpha=0$) and the equal-width method ($\alpha=5$), it is shown that by properly choosing the calibrating function $\gamma$, it is possible to find a more proper bucketing over traditional equal-frequency and equal-width methods.
\end{itemize}

\section{Experiments} 
\label{sec: exp}
In this section, we evaluate our CREAD framework on benchmark recommendation datasets, including two public recommendation datasets (CIKM\footnote{https://competitions.codalab.org/competitions/11161} and KuaiRec \cite{gao2022kuairec}) and a realistic industrial dataset (Industrial). In addition, we perform ablation studies to validate the design of the proposed framework, the training strategy, and the selection of the hyperparameters. 
We refer to Appendix \ref{app: exp} for the details of the datasets, model architectures, baselines, more ablation studies, and training details. 

\subsection{Experimental Settings}
\subsubsection{\label{sec:datasets}Datasets}
To make a fair comparison, we adopt the offline experiments on industrial datasets and two public datasets for reproducibility. 
The large-scale industrial dataset (referred to as \emph{Industrial}) is collected from a real-world streaming short-video platform. The dataset encompasses roughly $6$ billion impressions and over $50$ million users each day. We accumulated users' interaction logs for $14$ days, and subsequently use the next day's information for evaluation. 
Moreover, to improve reproducibility, we adopt  public benchmarks dataset KuaiRec \cite{gao2022kuairec} and  CIKM.  The KuaiRec dataset is  encompasses a total of $7,176$ users, $10,728$ distinct items, and a substantial count of $12,530,806$ impressions, while the evaluation dataset 
comprises $1,411$ users, $3,327$ items, and a collection of $4,676,570$ impressions. And the CIKM
dataset contains $87,934$ users, $122,994$ items and $1,235,381$ impressions.

\subsubsection{Evaluation Metrics} 
We follow previous works \cite{zhan2022deconfounding} and adopt the following metrics to measure watch time prediction quantitatively:
\begin{itemize}
\item \textbf{MAE (Mean Absolute Error)} is measured as the average of the absolute error between the watch time prediction $\hat{y}_i$ and the ground truth $y_i$, which is 
$\frac{1}{N}\sum_{i=1}^{N}|y_i-\hat{y}_i|.$

\item \textbf{XAUC \cite{zhan2022deconfounding}} is measured by randomly sampling instance pairs and checking whether their relative order is consistent with the ground truth. 


\end{itemize}

\subsubsection{Baselines\label{sec:baselines}} 



Here, we compare our method with previous methods for the watch time prediction task. The reference methods can be classified as traditional machine learning approaches and neural network approaches. The former includes Value Regression (VR), Weighted Logistic Regression (WLR) \cite{covington2016deep}, and Ordinal Regression (OR), while the latter includes the state-of-the-art model D2Q \cite{zhan2022deconfounding}. 
In particular, VR and WLR are continuous regression approaches (denoted as Cont.), while OR and D2Q break the target range into discrete buckets (denoted as Disc.). 
Since the authors did not provide an official code for watch time prediction, we implement them according to their corresponding papers, tune the number of buckets (if applicable) and report the best results. 

\begin{table}[t]

\centering
\begin{adjustbox}{width=.5\textwidth}

\begin{tabular}{ccccccc} 

\toprule
\multirow{2}{*} {Data} &Type
&\multicolumn{2}{c}{Cont.} &\multicolumn{3}{c}{Disc.} \\ \cmidrule(r){2-2} \cmidrule(r){3-4} \cmidrule(r){5-7}
&Method & VR & WLR & OR & D2Q & CREAD \\
\midrule
\multirow{2}{*}{Kuai.}&XAUC &0.5333 &0.5483&0.5883 & 0.5677 & \textbf{0.6009}  \\
&MAE &3.6992 &3.5722&3.2946 & 3.3090 & \textbf{3.2150}\\
\midrule
\multirow{2}{*}{CIKM}&XAUC & 0.6058    & 0.6310	    & 0.6313   & 0.6568  & \textbf{0.6671} \\
&MAE       & 2.3780	& 1.9785   & 1.9157	 & 2.5499	 & \textbf{1.7928} \\ 
\midrule
\multirow{2}{*}{Indust.}&XAUC &0.6324 &0.6323 &0.6344 &0.6423 & \textbf{0.6459}\\
&MAE &20.3672 &20.2740 & 20.1263 & 19.0740 & \textbf{18.6716}\\

\bottomrule
\end{tabular}
\end{adjustbox}
\caption{Watch time prediction results on KuaiRec, CIKM, and Industrial datasets: XAUC and MAE. 
}
\label{tab:sota}
\end{table}

\subsection{Comparison with State of the Art
\label{sec:sota_compare}}
In this section, we compare our method with current state-of-the-art methods for watch time prediction and report the MAE and XAUC metric in Table \ref{tab:sota}. 
We can see that the discrete approaches outperform the continuous regression approaches, indicating that the regression-by-classification framework is superior to direct regression in dealing with imbalanced data. Specifically, the discrete methods outperform continuous methods by at least $0.2632$ in MAE and $0.0194$ in XAUC on KuaiRec dataset. 
Besides, our CREAD outperforms all the discretization methods on these datasets with both metrics. Notably, CREAD outperforms the best (discretization) baseline D2Q by $5.85$\% on KuaiRec with XAUC, and $2.11$\% on Industrial with MAE. 


\begin{table*}[t]
\centering
\begin{adjustbox}{width=0.9\textwidth}

\begin{tabular}{lcccccccc}
\toprule
Metrics & $\textit{Equal Frequency}$  & $\beta = 0.5$ & $\beta = 1.0$ & $\beta = 3.0$ & $\beta = 10.0$ & $\textit{Equal Width}$ \\
\midrule
width(s) of tail bucket  &981.93  & 858.69 & 377.95 & 315.21 & 34.74 & 33.0 \\ \midrule
sample (\%) of tail bucket & 3.3\% & 0.1437\% & 0.0101\% & 0.0077\% & 0.0076\% & 0.0011\% \\ \midrule
XAUC ($\uparrow$) & 0.5962 & 0.5977 & 0.5992 & \textbf{0.6009} & 0.6001 & 0.5982 \\ \midrule

MAE ($\downarrow$) & 3.2477& 3.2419 & 3.2451 & \textbf{3.2150} & 3.2335 &  3.2492 \\
\bottomrule
\end{tabular}
\end{adjustbox}
\caption{Comparison on XAUC, MAE with different choices of label discretization on real-world data}
\label{tab:table1}
\end{table*}

\subsection{Ablation Studies}
We further conduct an ablation study on the proposed adaptive discretization, the ordinal regularization, and choices of the restoration function. 

\subsubsection{Effectiveness of Adaptive Discretization\label{sec:ablation}}

In Section \ref{sec:error-adaptive-discretization}, we explain the relation of our approach to the commonly used equal-width and equal-frequency discretization methods, as well as how our EAD balances them. To empirically verify this, we conduct an ablation study on KuaiRec by altering the discretization method and keeping other settings fixed. Specifically, we compare the equal-width discretization, equal-frequency discretization, and our adaptive discretization when $\beta \in \{0.5, 1.0, 3.0, 10.0\}$. 
As shown in Table \ref{tab:table1}, when we increase $\beta$ from $0.5$ to $10$, the width of the tail bucket, \ie the maximal bucket width, decreases monotonically from $858.69$s to $34.74$s, always less than the width of the equal-frequency discretization, and more than that of the equal-width discretization.
The same as the largest bucket width, the sample percentile of the tail bucket also reduces monotonically as $\beta$ grows. 
Moreover, the XAUC and MAE metrics are consistent, and both have a peak performance with $\beta=3.0$. Therefore, it is necessary to choose an appropriate $\beta$ that adapts to the realistic distribution. We adopt $\beta=3.0$ in all other experiments. 
Nevertheless, note that our EAD with any $\beta$ (the worst performance is with $\beta=0.5$) yields superior performance than equal-width discretization ($-0.0073$ in MAE) and equal-frequency discretization ($-0.0058$ in MAE). 
Finally, we observe that equal-frequency discretization achieves better performance ($-0.0015$ in MAE) than equal-width discretization when dealing with long-tailed data. 

\subsubsection{Effectiveness of Ordinal Regularization} 
\label{sec:loss_contraint_ablation}

\begin{figure}
\centering
\includegraphics[width=1.0\columnwidth]{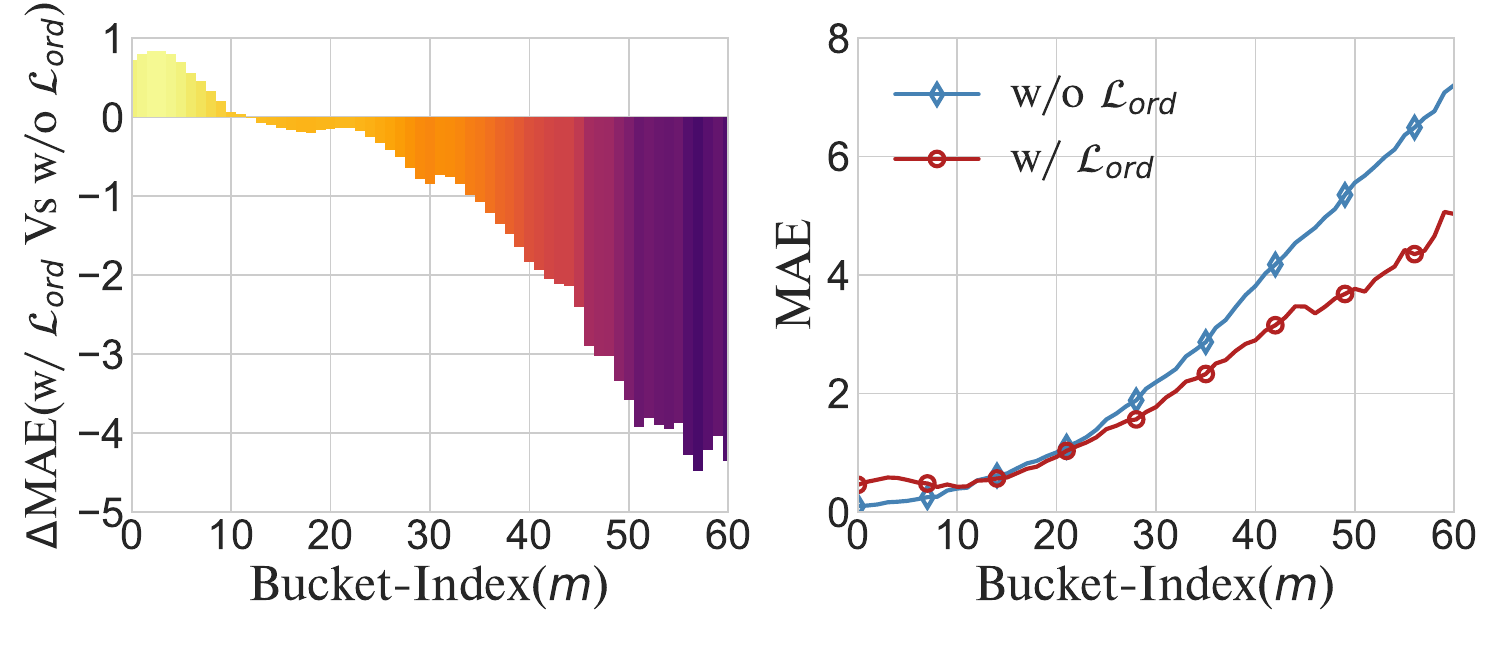}
\caption{Comparison between loss w/o and w/ ${\gL_{ord}}$ in MAE score with respect to bucket index.}
\label{fig:error_constraint_contribution_line}
\end{figure}
We verify the effectiveness of the ordinal regularization $\gL_{\text{ord}}$ introduced in Sec. \ref{sec:lossfunc} by training a model without it on KuaiRec. Experimental results show that ordinal regularization effectively improves performance by $0.0132$ in MAE, and $0.33$ in XAUC. 
Moreover, we show the performance change relative to our CREAD (denoted as $\Delta \text{MAE}$) as a function of the bucket index $m$ and report the difference in Figure \ref{fig:error_constraint_contribution_line},
from which we can draw several conclusions. 
First, whether with the ordinal regularization, MAE increases along the bucket index $m$, which is consistent with the long-tail distribution shown in Figure \ref{fig:watchtime_distribution}. 
Second, we observe that the ordinal regularization significantly reduces the error in the tail buckets. This is mainly because our EAD significantly reduces MAE in the tail buckets (Please refer to Appendix \ref{sec:exp_abl}), and our 
ordinal regularization further strengthens the effect. 
Finally, we attribute the performance degradation within the first $12$ buckets as a sacrifice to the overall generality.

\subsubsection{Influence of the Bucket Number}

\begin{figure}[t]
\centering
\includegraphics[width=0.7\columnwidth]{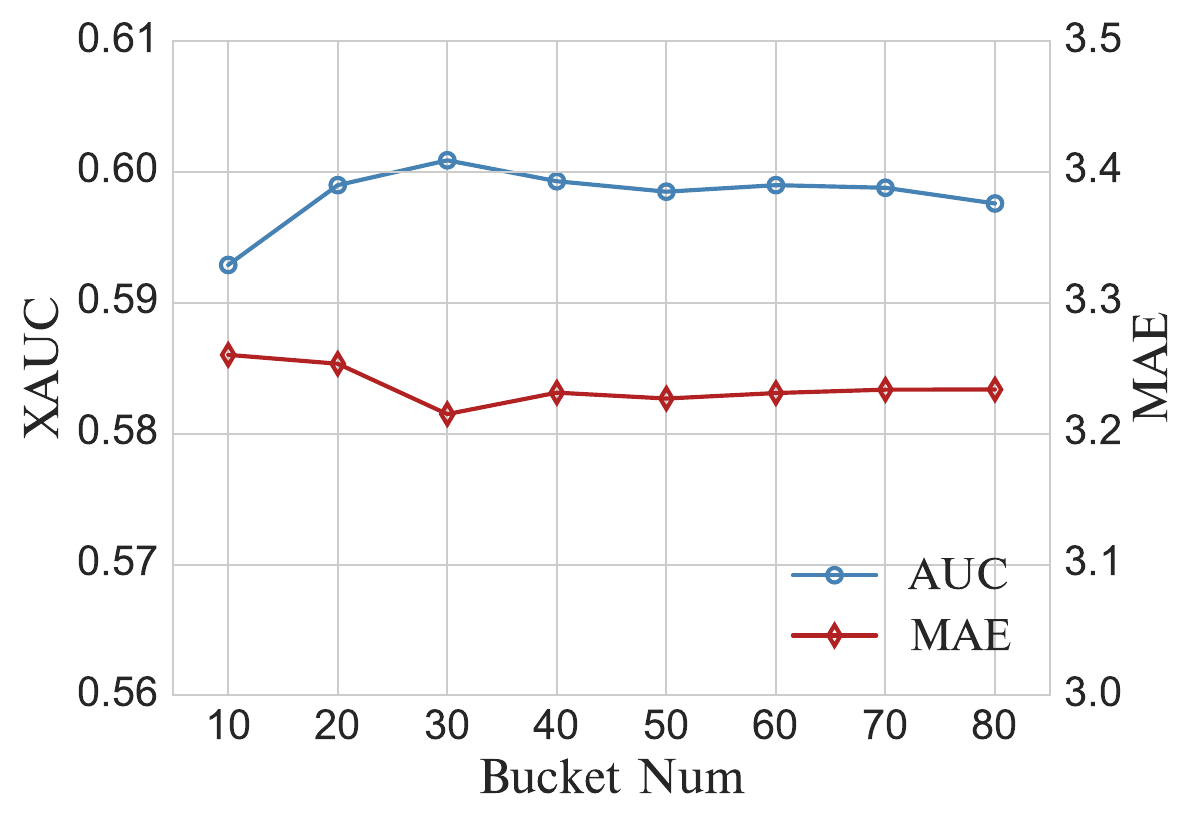}
\caption{ 
AUC and MAE of CREAD with different buckets. }
\label{fig:AUC-MAE-with-different-buckets}
\end{figure}

We investigate the influence of the bucket number $M$ in the discretization step. As shown in Figure \ref{fig:AUC-MAE-with-different-buckets}, given the trade-off between the MAE
and XAUC, we adopt $M=30$ in all other experiments.



\subsection{Online A/B Testing}

\begin{table}
\centering
\begin{adjustbox}{width=.5\textwidth}
\begin{tabular}{lccccc}
\toprule
\multirow{2}{*}{Days}
&Main Metric. & \multicolumn{4}{c}{Constraint Metrics} \\\cmidrule(r){2-2} \cmidrule(r){3-6}
&Watch-Time & Like&Share&Follow&Comment \\
\midrule
D1& +0.229\% &+0.132\% & -0.334\% & +0.102\% & +1.186\% \\
D2& +0.275\% &+0.230\% & -0.334\% & +0.364\% & -1.143\%  \\
D3& +0.273\% &+0.377\% & -0.112\% & +0.212\% & +2.337\% \\
D4& +0.327\% &+0.002\% & +0.221\% & +0.310\% & +1.781\% \\
D5& +0.353\% &-0.150\% & +0.157\% & +0.392\% & -0.782\%  \\
\bottomrule
\end{tabular}
\end{adjustbox}
\caption{Online A/B results.}
\label{tab:onlineab}
\end{table}

Besides offline experiments, we also conduct an online A/B test on Kwai App, a streaming short-video platform with more than $50$ million users each day. We attribute $20$\% traffic to our CREAD and the baseline D2Q \cite{zhan2022deconfounding}, respectively. 
We launch the experiments on the real-time system for $5$ days and report the results in Table \ref{tab:onlineab}. 
As we can see, 
the treatment group experienced a noteworthy increase in watch time (referred to as \emph{main metric}) by $0.291$\%. This outcome is particularly striking since production algorithms typically yield an average improvement of only $0.1$\% to $0.2$\%. Furthermore, we evaluate multiple metrics (referred to as \emph{main metric}) commonly used in actual production systems,
including the cumulative counts of \textit{like} (clicking the like button), \textit{follow} (subscribing to the video creator), \textit{share} (disseminating the video among friends), and \textit{comment} (providing textual remarks). 
The fluctuation of these metrics is not statistically significant. 




\section{Conclusion}
\label{sec: conclusion}
In order to predict user watch time in large-scale recommender systems, we propose a novel discretization framework, called CREAD, to learn from the skewed distribution and ordinal information of watch time. We analyze the error bounds introduced by discretization and propose a novel discretization approach to balance the learning error and the restoration error. Moreover, we also propose an ordinal regularization to capture the sequential prior between the discrete buckets. For offline and online experiments, we demonstrate that our model significantly improves the recommendation performance. 

\onecolumn
\newpage
\appendix
\section{Proofs}
\label{sec: proof}

\subsection{Proof of Lemma \ref{lemma:error-decomposition}}
\begin{proof}
We do the following decomposition:
\begin{equation}
    \hat{y} - y = D_p + D_w + D_b + D_y
\end{equation}
where
\begin{align}
 D_p =&   \sum_m\left(\hat{p}_m(\vx) - p_m(\vx)\right)\hat{w}_m \\ 
 D_w =& \sum_mp_m(\vx)\left(\hat{w}_m - w_m\right) \\
 D_b =&\sum_mp_m(\vx)\left(w_m - v_m(\vx)\right) \\
 D_y =& \sum_mp_m(\vx)v_m(\vx)-y
\end{align}
Then we have
\begin{align}
    V_p &= \bbE D_p^2 \\
    V_w &= \bbE D_w^2 \\
    V_b &= \bbE D_b^2 \\
    V_y &= \bbE D_y^2
\end{align}

We prove the lemma according to the following fact: for a random variable $z$, if $\bbE z=0$, then $\bbE \left(z + a\right)^2 = \bbE z^2 + a^2$. Therefore, if we can prove that $\bbE_{\hat{p}} D_p = \bbE_{\hat{w}} D_w = \bbE_{\vx} D_b = 0$, then we will have
\begin{equation}
\begin{aligned}
    \bbE\left(\hat{y}-y\right)^2 &= \bbE\left(D_p+D_w+D_b+D_y\right)^2 \\
    &=\bbE_{\vx,y,\hat{w}}\left[\bbE_{\hat{p}}D_p^2+(D_w+D_b+D_y)^2\right] \\
    &=\bbE_{\vx,y,\hat{w},\hat{p}}D_p^2 + \bbE_{\vx,y}\left[\bbE_{\hat{w}}D_w^2 + (D_b+D_y)^2\right] \\
    &= \bbE_{\vx,y,\hat{w},\hat{p}}D_p^2 + \bbE_{\vx,y,\hat{w}}D_w^2 +\bbE_{\vx,y}\left(D_b^2+D_y^2\right)  \\
    &= \bbE_{\vx,\hat{w},\hat{p}}D_p^2 + \bbE_{\vx,\hat{w}}D_w^2 + \bbE_{\vx}D_b^2 + \bbE_{\vx,y}D_y^2 \\
    &= V_p + V_w + V_b + V_y.
\end{aligned}
\end{equation}
Therefore, it suffices to prove $\bbE_{\hat{p}} D_p = \bbE_{\hat{w}} D_w = \bbE_{\vx} D_b = 0$.

Since $\hat{p}_m$ is the unbiased estimation of $p_m$ by assumption, i.e., $\bbE_{\hat{p}}\left(\hat{p}_m(\vx) - p_m(\vx)\right) = 0$, we have:
\begin{equation}
\bbE_{\hat{p}} D_p = \sum_m\hat{w}_m\bbE_{\hat{p}}\left(\hat{p}_m(\vx) - p_m(\vx)\right) = 0
\end{equation}

Similarly, according to the assumption that $\hat{w}_m$ is the unbiased estimation of $w_m$, i.e., $\bbE_{\hat{w}}\left(\hat{w}_m - w_m\right) = 0$, we have
\begin{equation}
    \bbE_{\hat{w}} D_w = \sum_mp_m(\vx)\bbE_{\hat{w}}\left(\hat{w}_m - w_m\right) = 0
\end{equation}

According to the definition that $w_m = \bbE_{\vx}v_m(\vx)$, we have
\begin{equation}
\begin{aligned}
\bbE_{\vx}p_m(\vx)w_m = P(y\in d_m)\bbE_\vx\bbE\left[y|\vx,y\in d_m\right] = P(y\in d_m)\bbE\left[y|y\in d_m\right] = \bbE\left[y\cdot\bone(y\in d_m)\right].
\end{aligned}
\end{equation}

Also,
\begin{equation}
    \bbE_\vx p_m(\vx)v_m(\vx) = \bbE_\vx P(y\in d_m|\vx)\bbE\left[y|\vx,y\in d_m\right] = \bbE_\vx\bbE\left[y\cdot \bone(y\in d_m)|\vx\right] = \bbE\left[y\cdot \bone(y\in d_m)\right]
\end{equation}
Therefore, we have $\bbE_{\vx} D_p = \sum_m\left[\bbE_{\vx} p_m(\vx)w_m - \bbE_{\vx} p_m(\vx)v_m(\vx)\right] =0$, which finishes the proof.
\end{proof}

\subsection{Proof of Theorem \ref{thm:error-bound}}
We first provide some lemma on the variance of $\hat{p}(\vx)$ and $\hat{w}_m$.
\begin{lemma} \label{lemma:p-m}
    Assume the input set $\mathcal{X}$ is finite, then we have
    \begin{equation}
        \bbE_{\hat{p}} \left[\hat{p}_m(\vx) - p_m(\vx)\right]^2 = \frac{1}{N_\vx}p_m(\vx)\left[1-p_m(\vx)\right]
    \end{equation}
    where $N_\vx$ is the number of samples with input $\vx$.
\end{lemma}
\begin{proof}
Since the input set $\mathcal{X}$ is finite, a maximum likelihood estimation of $p(\vx)$ is $\hat{p}_m(\vx) = N_{\vx,y\in d_m} / N_\vx$, i.e., the frequency of $y$ falling into $d_m$ under $N_\vx$ samples. Then, $N_{\vx,y\in d_m}$ holds a binomial distribution 
 $B(N_\vx, p_m(\vx)$, of which the expectation is $N_\vx p_m(\vx)$ and variance is $N_\vx p(\vx)\left[1-p(\vx)\right]$. By dividing this formula by $N_\vx$, we obtain that the expectation of $\hat{p}_m$ is $p_m(\vx)$, and the variance is $p_m(\vx)\left[1-p_m(\vx)\right]$, which finishes the proof.
\end{proof}

\begin{lemma} \label{lemma:w-m}
    The estimation of the representing value $w_m$, i.e., $\hat{w}$, has a variance term:
    \begin{equation}
        \bbE_{\hat{w}} \left(\hat{w}_m-w_m\right)^2 = \frac{\text{var}\left[y|y\in d_m\right]}{NP(y\in d_m)}
    \end{equation}
\end{lemma}

\begin{proof}
    Note that $\hat{w}_m$ is the estimate of the expectation of $y$ conditioned on $y\in d_m$. A maximum likelihood estimation gives $\hat{w}_m = \sum_{y\in d_m}y / N(y\in d_m)$, i.e., an empirical average value of the watch time $y$ in the interval $d_m$. Since the variance of one sample in $d_m$ is $\text{var}\left[y|y\in d_m\right]$, given the \textit{i.i.d.} assumption, the variance of $\hat{w}_m$ should be $1/\left[NP(y\in d_m)\right]$ the variance of one sample in $d_m$, which finish the proof.
\end{proof}

\begin{proof}[Proof of the Upper Bound of $V_p$ in Theorem \ref{thm:error-bound}]
According to the Cauchy's inequality,
\begin{equation}
    V_p \leq M\sum_m\bbE_{\vx,\hat{w},\hat{p}}\left[\left(\hat{p}_m(\vx)-p_m(\vx)\right)\hat{w}_m\right] ^2.
\end{equation}
Then, note that $\left|\hat{w}_m\right|\leq t_m$, we have
\begin{equation}
\begin{aligned}
    V_p &\leq M\sum_m\bbE_{\vx,\hat{p}}\left(\hat{p}_m(\vx)-p_m(\vx)\right)^2t_m^2 \\
    &= M\sum_m\bbE_\vx \frac{1}{N_\vx}p_m(\vx)(1-p_m(\vx))t_m^2 \\
    &\leq M\sum_m\bbE_\vx \frac{1}{N_\vx}p_m(\vx)t_m^2 \\
    &= M\sum_m\sum_\vx \mu(\vx)\frac{1}{N_\vx}p_m(\vx)t_m^2
\end{aligned}
\end{equation}
Note that we have used Lemma \ref{lemma:p-m} to calculate $\bbE_{\hat{p}}\left(\hat{p}_m(\vx)-p_m(\vx)\right)^2$. Then, according to the fact that the probability of $\vx$, i.e., $\mu(\vx)$, equals $N_\vx / N$, we have
\begin{equation}
\begin{aligned}
    V_p &\leq \frac{M}{N}\sum_mt_m^2\sum_\vx p_m(\vx) \\
    &=\frac{M\left|\mathcal{X}\right|}{N}\sum_m P(y\in d_m)t_m^2
\end{aligned}
\end{equation}
Note that $\sum_m P(y\in d_m)t_m^2$ is an approximation of $\bbE_{y\in\Psi}y^2$ by discretization. Then, if we assume $\Psi$ to be bounded second differentiable (based on the assumption of Theorem \ref{thm:error-bound}), then there must exist a constant $C_p$ that $\sum_m P(y\in d_m)t_m^2 \leq C_p \bbE_{y\in\Psi}y^2$ \cite{hildebrand1987introduction}. Therefore,
\begin{equation}
    V_p \leq \frac{C_pM\left|\mathcal{X}\right|}{N}\bbE_{y\in\Psi}y^2
\end{equation}
which finishes the proof of the error bound of $V_p$.

\end{proof}
\begin{proof}[Proof of the Upper Bound of $V_w$ in Theorem \ref{thm:error-bound}]
According to the Cauchy's inequality, 
\begin{equation} \label{eq:V_w-1}
    V_w \leq \sum_m  p_m^2 \cdot \sum_m\bbE(\hat{w}_m-w_m)^2
\end{equation}
There are two multipliers on the right of the inequality. Consider the first term, and note that $p_m$ is the probability of the watch time $y$ falling into the interval $d_m$, i.e., $\Psi(t_m)-\Psi(t_{m-1})$, then we have
\begin{equation} \label{eq:sum-p-m}
    \sum_m p_m^2 = \sum_m \left[\Psi(t_m)-\Psi(t_{m-1})\right]^2
\end{equation}
Then, consider the second term in \eqref{eq:V_w-1}, and according to Lemma \ref{lemma:w-m}, we have  
\begin{equation}
\bbE(\hat{w}_m-w_m)^2 = \frac{\text{var}\left[y|y\in d_m\right]}{N\left[\Psi(t_m)-\Psi(t_{m-1})\right]}
\end{equation}
Moreover, according to the bounded second differentiability of $\mu(\vx, y)$, we have $\text{var}\left[y|y\in d_m\right] \leq C_w\left(t_m-t_{m-1}\right)^2$, where $C_w$ is a certain constant. Therefore,
\begin{equation}
    V_w \leq \frac{C_w}{N}\sum_m \left[\Psi(t_m)-\Psi(t_{m-1})\right]^2\cdot  \sum_m \frac{\left(t_m-t_{m-1}\right)^2}{\left[\Psi(t_m)-\Psi(t_{m-1})\right]}
\end{equation}
which finishes the proof of the error bound of $V_w$.
\end{proof}
\begin{proof}[Proof of the Upper Bound of $V_b$ in Theorem \ref{thm:error-bound}]
Also, use Cauchy's inequality,
\begin{equation}
    V_b \leq \sum_m p_m^2 \cdot \sum_m\bbE_\vx\left(w_m-v_m(\vx)\right)^2
\end{equation}
the term $\sum_m p_m^2$ is the same as \eqref{eq:sum-p-m}, hence we only consider the second term $\sum_m\bbE_\vx\left(w_m-v_m(\vx)\right)^2$, which is the variance of $v_m(\vx)$, i.e. $\text{var}\left[y|y\in d_m\right]$. Therefore we have $\sum_m\bbE_\vx\left(w_m-v_m(\vx)\right)^2 \leq C_b (t_m-t_{m-1})^2$, and
\begin{equation}
    V_b \leq C_b \sum_m \left[\Psi(t_m)-\Psi(t_{m-1})\right]^2\cdot  \sum_m \left(t_m-t_{m-1}\right)^2
\end{equation}
which finishes the proof.
\end{proof}
\section{More on Experiments}
\label{app: exp}
\subsection{Datasets}
The KuaiRec dataset \cite{gao2022kuairec} is sourced from a streaming short-video platform called Kuaishou with users' interaction logs. 
The training dataset encompasses a total of $7,176$ users, $10,728$ distinct items, and a substantial count of $12,530,806$ impressions, while the evaluation dataset 
comprises $1,411$ users, $3,327$ items, and a collection of $4,676,570$ impressions. 
In terms of viewing behavior, the average watch time per video in the training data amounts to $9.02$ seconds, with a median of approximately $7.27$ seconds. Notably, the training dataset features a maximum watch time of $999.63$ seconds, following a long-tailed distribution pattern.

The CIKM dataset is collected by extracting users sessions from an e-commerce search engine logs with time spending on each item. The training dataset contains $75,292$ users,
$118,164$ distinct items and $1,111,851$ impressions, while the evaluation dataset contains $13,292$ users,
$37,154$ items and $123,529$ impressions.

The industrial dataset accumulated through $14$ days, where each day comprises roughly $6$ billion training samples. We subsequently use the next day for model evaluations. The dataset's videos exhibit an average duration of 26.6 seconds, with a median duration of 13.9 seconds, aligning cohesively with the observed watch time distribution.
\subsection{Model Structure}
\label{app:model_structure}
It is important to emphasize that the CREAD framework exhibits compatibility with various model structures. Accordingly, we adopt a multi-layer perceptron (MLP) network as the foundational architectural choice. To ensure fair comparisons, all baseline methods share the same architecture 
. A comprehensive depiction of the model architecture is available in Appendix \ref{app:model_structure}.

\begin{figure}[t]
\centering
\includegraphics[width=0.6\columnwidth]{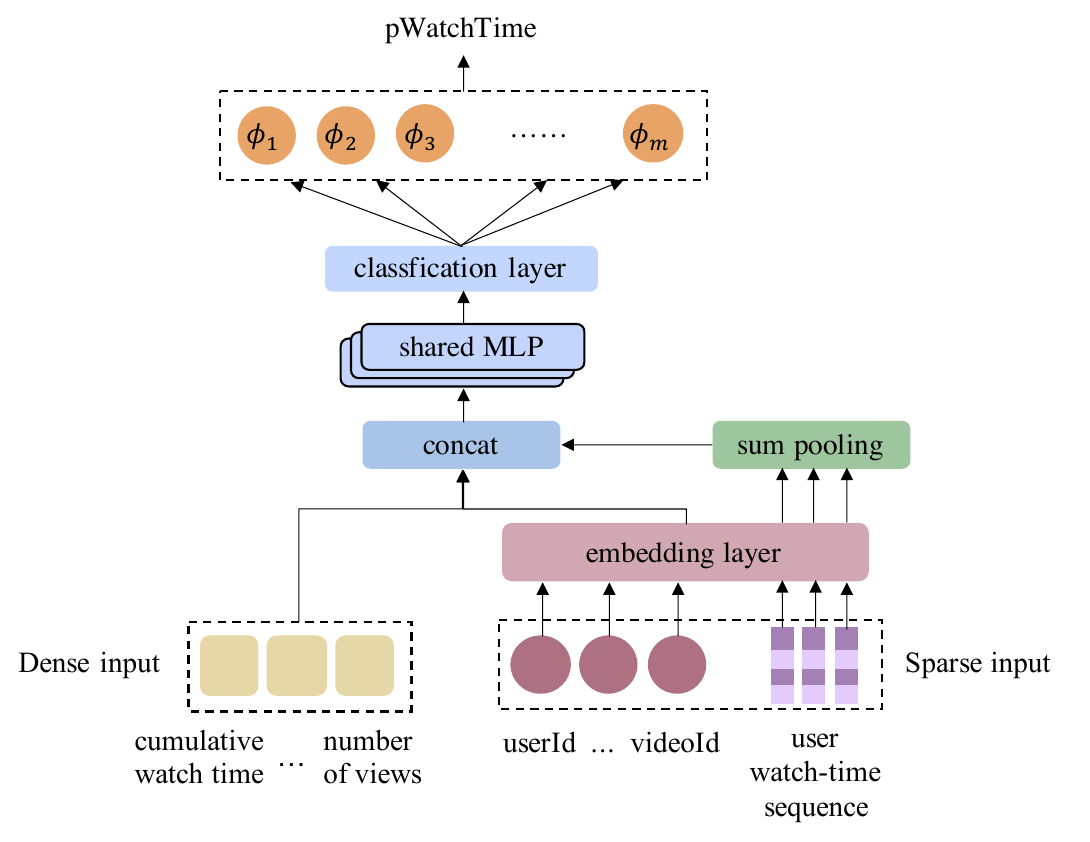}
\caption{Model architecture. “Dense input” refers to statistical features (such as \textit{cumulative watch time
} and \textit{number of views}) of the video. “Sparse input” refers to ID features (such as \textit{device ID}, \textit{user ID}, and \textit{video ID}), categorical features (such as \textit{user gender} and \textit{video category}), and user sequence behavior (such as \textit{recent video play list} and \textit{recent author play list}).}
\label{fig:model_structure}
\end{figure}

\subsection{Baselines}

\begin{itemize}
\item \textbf{VR (Value Regression)}. This approach directly predicts the absolute value of watch time, where the learned function's accuracy is evaluated by mean squared error. 
\item \textbf{WLR (Weighted Logistic Regression)}\cite{covington2016deep}. This method frames watch time regression as a binary classification problem, designating impressed and clicked videos as positive samples, and impressed yet unclicked ones as negative samples, with the losses of positive samples incorporating watch time-based weighting. The acquired odds serve as an approximate representation of watch time, effectively integrating watch time as learned odds within the logistic regression model. In this schema, positive samples carry watch time-based weightage, while negative samples maintain unit weight.
\item \textbf{OR (Ordinal Regression)}. To implement Ordinal Regression (OR) for watch time predictions, we begin by discretizing continuous watch time values into $K$ meaningful ordinal categories. Each category is then associated with a classifier, predicting whether the forecast exceeds the discretized threshold. During prediction, probabilities or odds for each category are computed and subsequently translated to the continuous scale. Through multiple trials, we determine the optimal discretization value of $K=80$, which yields superior performance.

\item \textbf{D2Q (Duration-Deconfounded Quantile-based)}\cite{zhan2022deconfounding}. The D2Q method is designed to address the issue of duration bias in predicting video watch time. It accomplishes this by dividing data into $10000$ watch time percentiles based on the video duration and fitting a regression model on these percentiles to approximate watch time. This process culminates in the derivation of predicted watch time through the judicious mapping of forecasted quantiles onto the watch time domain. Our empirical investigations show that a duration group number set at $30$ yields the optimal balance between granularity and predictive precision.
\end{itemize}

\subsection{Hyperparameter Analysis}
Unless otherwise stated, we adopt the following training protocol.
The hyperparameters $\lambda_{\text{ce}}, \lambda_{\text{ord}},$ and $\lambda_{\text{restore}}$ in equation \eqref{eq:final_loss} are deliberately set to $100.0$,  $10.0$ and $1.0$, respectively, without adding additional fine-tuning. This choice serves the purpose of balancing the loss magnitudes across these specific loss components.
We train for $10$ epochs in total with a batch size of $1024$, using the Adam optimizer \cite{kingma2014adam} with default parameters $\beta_1 = 0.9$ and $\beta_2 = 0.999$ to minimize the loss function.






\subsection{More Ablation Studies}
\label{sec:exp_abl}
\subsubsection{Choice of the restoration error}
\begin{figure}[t]
\centering
\includegraphics[width=0.8\columnwidth]{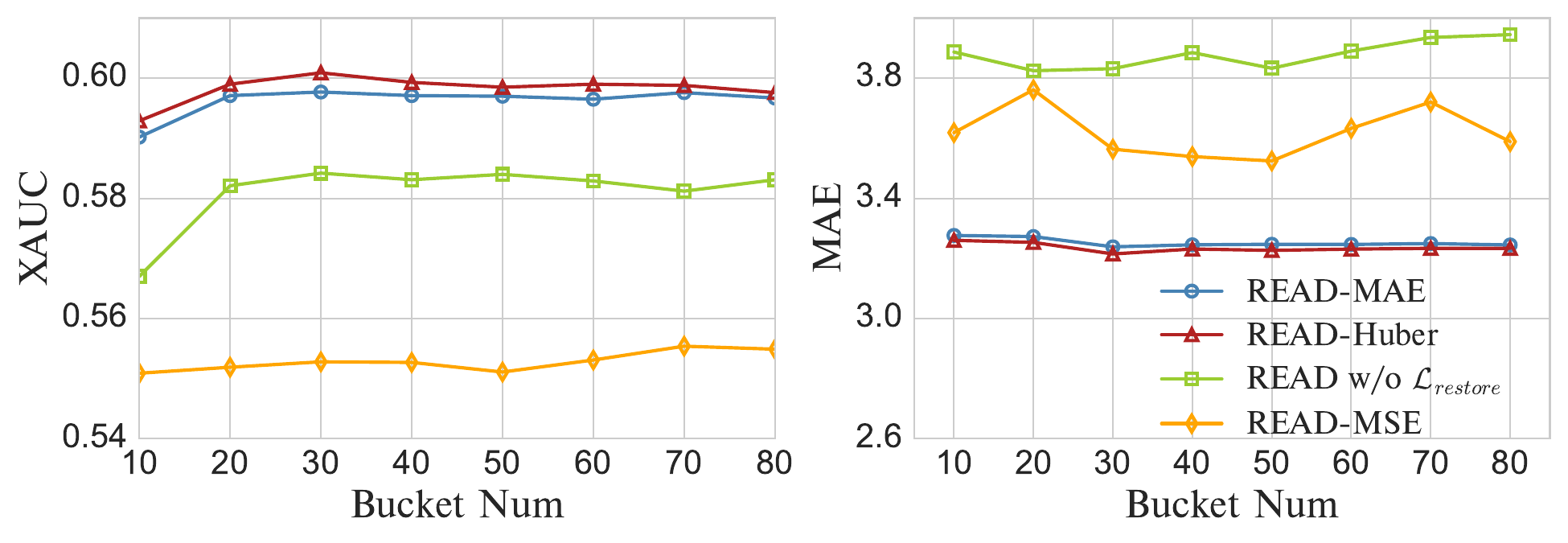}
\caption{
AUC and MAE of CREAD with different Losses.}
\label{fig:AUC-MAE-with-different-loss}
\end{figure}
We study the influences of imposing accuracy loss $\gL_{\text{restore}}$ on final prediction values. We consider the following form of $\gL_{\text{restore}}$, of which the Huber loss performs the best: 
\begin{itemize}
\item {\bfseries Mean Squared Error (MSE) Loss:} This arithmetic mean-unbiased loss function is commonly used in regression learning, measuring the sum of squared differences between actual and predicted values. However, it is highly susceptible to outliers. Since the distribution of watch time is discernibly right-skewed, using MSE loss can overestimate the impact of videos with extremely large value of watch time, magnify the gradient during training, and degrade model stability.
\item {\bfseries MAE Loss:} This loss measures the absolute differences between true and predicted values and is less sensitive to outliers compared with MSE. However, it also neglects prediction accuracy for tail samples.
\item {\bfseries Huber Loss.} The Huber loss function is a combination of the MSE and MAE loss, which effectively mitigates the shortcomings of its predecessors. The Huber loss function is defined as follows: 
\begin{align}
    \gL_{restore} = 
    \left\{
    \begin{aligned}
        & \frac{1}{2} (T-\hat{T} )^2, & \text{if } |T-\hat{T}| < \delta, \\
        & \delta|T-\hat{T}| -\frac{1}{2} \delta^2, & \text{if } |T-\hat{T}|>\delta,
    \end{aligned}
    \right.
\end{align} 
where $ \delta $ is a hyperparameter that controls the split between the two sub-functions. Compared with MSE and MAE, the Huber loss function avoids excessive sensitivity to large errors and overcomes the corner issues of MAE when the error is relatively small.
\end{itemize}

We take a series of trials to test the performance of Huber loss, MAE loss, and MSE loss with sample partition size $(M = 10, 20, 30, 40, 50, 60, 70, 80) $. 
Figure \ref{fig:AUC-MAE-with-different-loss} presents the results. It is shown that by adding $\gL_{restore}$ we significantly improve the overall model performance and Huber loss outperforms the other losses.

\subsection{Online A/B Test}
The A/B experiments were constructed as follows: Incoming traffic was systematically divided into ten equitably sized buckets. Two of these buckets were attributed to the baseline condition, while the remaining two were earmarked for our proposed method. Given that our online experiment is executed on a video platform catering to more than $50$ million daily users, the findings garnered from analyzing a mere $20$\% of the total traffic stand as remarkably compelling. The models were trained using an online learning approach, employing streaming data derived from actual real-time streaming message systems, such as Apache Kafka. While watch time is an essential component of the ranking process in large-scale recommender systems, it is not the sole factor that must be considered. The ranking score depends on a comprehensive set of predicted values for various user behaviors. For the purposes of this paper, we have modified only the prediction of watch time, while leaving other prediction models unchanged, such as users' like rate and follow rate. 
\end{document}

%% file: math_commands.tex
\usepackage{amsmath,amsfonts,bm}

\def\eqref#1{Eq.~(\ref{#1})}

\def\1{\bm{1}}

\def\vx{{\bm{x}}}

\DeclareMathAlphabet{\mathsfit}{\encodingdefault}{\sfdefault}{m}{sl}
\SetMathAlphabet{\mathsfit}{bold}{\encodingdefault}{\sfdefault}{bx}{n}

\def\gD{{\mathcal{D}}}

\def\gL{{\mathcal{L}}}

\def\gY{{\mathcal{Y}}}

\renewcommand{\epsilon}{\varepsilon}

\newcommand{\bx}{\mathbf{x}}

\newcommand{\bbE}{\mathbb{E}}

\newcommand{\bbR}{\mathbb{R}}

\newcommand{\bone}{\mathbf{1}}

\def\ie{\textit{i.e.,~}}
\def\eg{\textit{e.g.,~}}